# Metrology Camera System of Prime Focus Spectrograph for Subaru Telescope

Shiang-Yu Wang*[a], Richard C. Y. Chou[a], Pin-Jie Huang[a], Hung-Hsu Ling[a], Jennifer Karr[a], Yin-Chang Chang[a], Yen-Shan Hu[a], Shu-Fu Hsu[a], Hsin-Yo Chen[a], James E. Gunn[b], Dan J. Reiley[c], Naoyuki Tamura[d], Naruhisa Takato[e], Atsushi Shimono[d]

[a]Academia Sinica, Institute of Astronomy and Astrophysics, P. O. Box 23-141, Taipei, Taiwan;
[b]Princeton University, Princeton, New Jersey, 08544, USA;
[c]California Institute of Technology, 1200 E California Blvd, Pasadena, CA 91125, USA;
[d]Kavli Institute for the Physics and Mathematics of the Universe (WPI),The University of Tokyo Institutes for Advanced Study, The University of Tokyo, Kashiwa, Chiba 277-8583, Japan;
[e]Subaru Telescope, National Astronomical Observatory of Japan, 650 North A'ohoku Place Hilo, HI 96720, U.S.A.

## ABSTRACT

The Prime Focus Spectrograph (PFS) is a new optical/near-infrared multi-fiber spectrograph designed for the prime focus of the 8.2m Subaru telescope. PFS will cover a 1.3 degree diameter field with 2394 fibers to complement the imaging capabilities of Hyper SuprimeCam. To retain high throughput, the final positioning accuracy between the fibers and observing targets of PFS is required to be less than 10 microns. The metrology camera system (MCS) serves as the optical encoder of the fiber motors for the configuring of fibers. MCS provides the fiber positions within a 5 microns error over the 45 cm focal plane. The information from MCS will be fed into the fiber positioner control system for the closed loop control.

MCS will be located at the Cassegrain focus of Subaru telescope in order to to cover the whole focal plane with one 50M pixel Canon CMOS camera. It is a 380mm Schmidt type telescope which generates a uniform spot size with a ~10 micron FWHM across the field for reasonable sampling of the point spread function. Carbon fiber tubes are used to provide a stable structure over the operating conditions without focus adjustments. The CMOS sensor can be read in 0.8s to reduce the overhead for the fiber configuration. The positions of all fibers can be obtained within 0.5s after the readout of the frame. This enables the overall fiber configuration to be less than 2 minutes. MCS will be installed inside a standard Subaru Cassgrain Box. All components that generate heat are located inside a glycol cooled cabinet to reduce the possible image motion due to heat. The optics and camera for MCS have been delivered and tested. The mechanical parts and supporting structure are ready as of spring 2016. The integration of MCS will start in the summer of 2016.

In this report, the performance of the MCS components, the alignment and testing procedure as well as the status of the PFS MCS will be presented.

## 1. INTRODUCTION

The Prime Focus Spectrograph (PFS) is a new prime focus optical/near-infrared multi-fiber spectrograph for the 8.2m Subaru telescope[1]. PFS will cover a 1.3 degree diameter field with 2394 fibers to complement the imaging capabilities of Hyper SuprimeCam[2]. To retain high throughput, the final positioning accuracy between the fibers and observing targets of PFS is required to be less than 10µm. The metrology camera serves as the optical encoder of the fiber motors for the configuring of fibers[3]. The metrology camera is designed to provide the fiber position information within 5µm error over the 45cm focal plane. The information from the metrology camera will be fed into the fiber positioner control system for the closed loop control.

The metrology camera will be located at the Cassegrain focus of Subaru telescope to cover the whole focal plane with one 50M pixel Canon CMOS sensor. To reduce high spatial frequency distortion of the wide field corrector (WFC), the aperture size of the metrology camera is set to be 380 mm which is the largest affordable aperture at the Cassegrain focus[4]. A Schmidt telescope type optical design was adapted to provide uniform image quality across the field with reasonable sampling of the point spread function (PSF). The mechanical design based on Invar and carbon fiber tubes provides stable structure over the temperature range under operating conditions. The CMOS sensor can be read in 0.8s to reduce the overhead for the fiber configuration[4]. The positions of all fibers can be obtained within 1s after the exposure is finished. This enables the overall fiber configuration to be less than 2 minutes.

* sywang@asiaa.sinica.edu.tw; phone 886 2 2366-5338; fax 886 2 2367-7849; www.asiaa.sinica.edu.tw

The PFS collaboration is led by the Kavli Institute for the Physics and Mathematics of the Universe and the University of Tokyo with international partners consisting of Academia Sinica, Institute of Astronomy and Astrophysics in Taiwan, Caltech/Jet Propulsion Laboratory and Princeton University/John Hopkins University in the USA, the Chinese PFS Participating Consortium in China, the Laboratoire d'Astrophysique de Marseille in France, the Max Planck Institute for Astronomy in Germany, the National Astronomical Observatory of Japan/Subaru Telescope, and the Universidade de São Paulo/Laboratório Nacional de Astrofísica in Brazil.

## 2. METROLOGY CAMERA FUNCTIONS AND COMPONENTS

The metrology camera will be installed inside a Cassegrain instrument box as for other Subaru Cassegrain instruments. Calibration of the image distortion and mapping to the focal plane through the WFC is achieved by back-lit fixed fiducial fibers and the home positions of scientific fibers. The home positions of the science fibers have been tested to provide good repeatability (~1µm) after rotational movements. The positions for the fiducial fibers and the home positions of the science fibers will be measured to a high repeatable accuracy during the integration and verification of the fiber system.

In the beginning of each COBRA configuration, the COBRAs will move to the home position to generate reference points across the field for distortion map confirmation. After that, the science fibers will move to new positions based on the chosen target to fiber assignment. The MCS will measure the location of each fiber position with the sensor coordinate system and send it to the PFI control system. Then, the fiber position errors are calculated with respect to the required locations for the fibers. This iteration will be repeated a number of times until the errors converge to the expected value. The back illuminator of fiducial and science fibers will be turned on throughout the iterations. In addition, MCS will also support two additional diagnostic functions during the commissioning and engineering phase. The first is to image the circular motion of the COBRA fibers to obtain the rotational center of the motors. During such an operation, a series of images will be taken with fibers stopping at several positions on the circular moving trajectories. MCS will calculate the center position of the trajectories with high precision. The second function is to generate fiber images when the instrument rotator is moving while COBRA stays still. With several exposures during the rotator movement, the metrology camera will capture the arcs generated by the fibers and estimate the offset and tilt between the rotator axis and the fibers focal plane center.

Some MCS components have been presented in the last SPIE proceeding[2,4]. The MCS development had the critical design review (CDR) in September 2015. In the following sections, we focus only on the details and status of the MCS optics and mechanical structure.

### 1.1. Camera optics

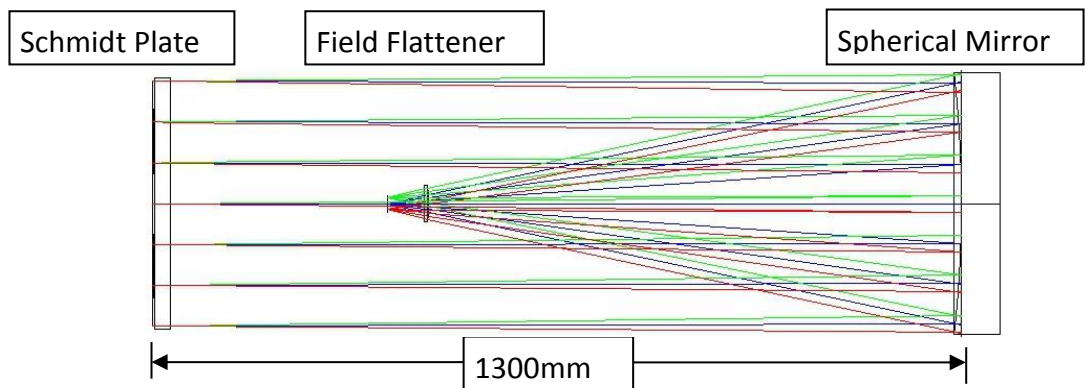

Figure 1. The optical layout of the metrology camera.

The optics of the metrology camera should be able to provide uniform optical performance and negligible image distortion across the field. Considering the size of the sensitive area of the detector used, a magnification factor of 0.038 was chosen for the metrology camera, which yields a focal length of 847mm. Considering the central hole size at the Cassegrain focus of Subaru telescope, the aperture size of the metrology camera is set to be 380mm. With a 380mm aperture, the beam deviation due to WFC can be reduced to around 2.5µm[4].

The metrology camera has fast optics (f/2.2) to accommodate the 380mm aperture size and a focal length of 847mm. The Schmidt reflector design enables a compact system to fit into the Subaru Cassegrain box with a height limit of 1750mm. The camera optics consists of a Schmidt plate, a spherical mirror and a field flattener. Figure 1 shows the optical layout of the design. In order to generate well sampled image spots, the Schmidt plate was deliberately designed to under correct the spherical aberration. The predicted spot size of the fiber tip extends about 9µm or ~3 pixels. The shape of the PSFs over the entire field is quite uniform and satisfies the centroid estimation requirement. We conducted tolerance analysis and figure 2 shows the simulated as-built PSF images at the field center and edge, which considers the obscurations in light path, potential manufacturing and alignment errors and optics deformations. The slightly oblate PSF is mainly due to the asymmetrical obscurations in light path; the FWHM of the PSF is ~10µm. The image distortion at the field edge is 0.01% which is much smaller than the distortion of WFC (2.5%).

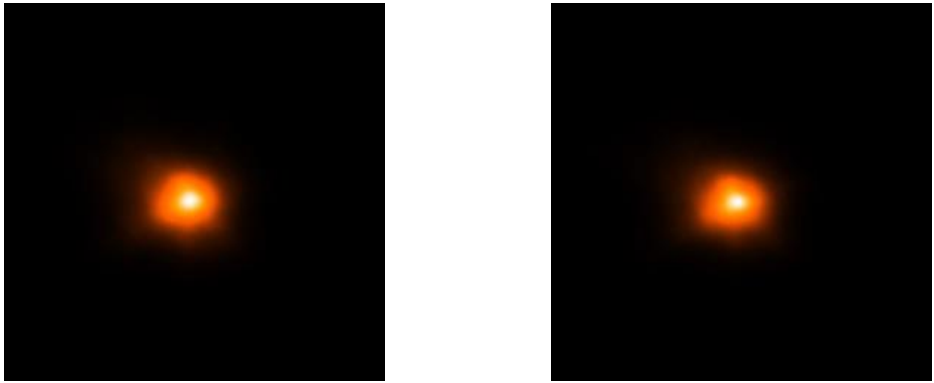

Figure 2. As-built PSF for field center (left) and field edge (right) of the PFS metrology camera after considering all degradation factors.

All the optics have been fabricated (see figure 3) and inspected to meet the design requirements. The Schmidt plate was fabricated by Nikon Corporation. The aspherical surface has R.M.S. surface sag error less than the required 0.09µm; the flat surface has curvature error less than six fringes and surface irregularity less than one fringe. The spherical mirror and the field flattener were fabricated by Control Optics Taiwan Inc. The curvature error is less than 0.1% of radius and the surface irregularity is better than the required 1 fringe (0.8 fringes). All the lenses are made of silica with anti-reflection coating to achieve a reflectance of less than 2% between 400 and 700 nm, allowing day time operation for system calibration. The spherical mirror was made of Zerodur with protected aluminum coating to minimize thermal expansion/contraction.

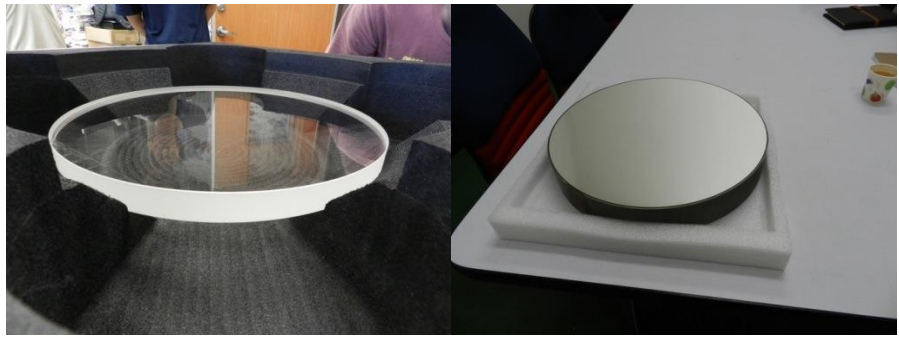

Figure 3. The left panel shows the Schmidt plate fabricated by Nikon Corporation. The right panel shows the spherical mirror fabricated by Control Optics Taiwan Inc.

## 1.2. Camera mechanics

The design principle for the mechanical structure is to reduce possible adjustments during observations. The main supporting structure is made of carbon fiber tubes to ensure very low thermal expansion and low deformation due to gravity. As a Subaru Cassegrain instrument, the mechanical structure for the metrology camera should meet the space and mounting scheme as for other Cass instruments. The height of the Cass box is 1750mm which sets a strong constraint on the optical design of the metrology camera. The total weight of the metrology camera is roughly 260kg, including the mechanical structures and optics. The mechanical structure for the metrology camera can be separated into two components: 1) the supporting and mounting structure for the optical elements and camera sensor module (see figure 4 left); 2) a box with mounting flange to match the telescope interface (figure 4 right). This provides the required interface to fit our system to the Cassgrain bonette of Subaru telescope so that the system can be positioned with a repeatable accuracy. The camera mounting structure is connected to the Cassegrain box through a top plate. The top plate also carries the counter weight and is fully sealed so that the heat generated from the metrology camera system will not flow up to the telescope optical path. Inside the Cassegrain box, the components which generate heat are located inside a sealed cabinet with glycol cooling to keep the temperature of the Cassegrain box stable.

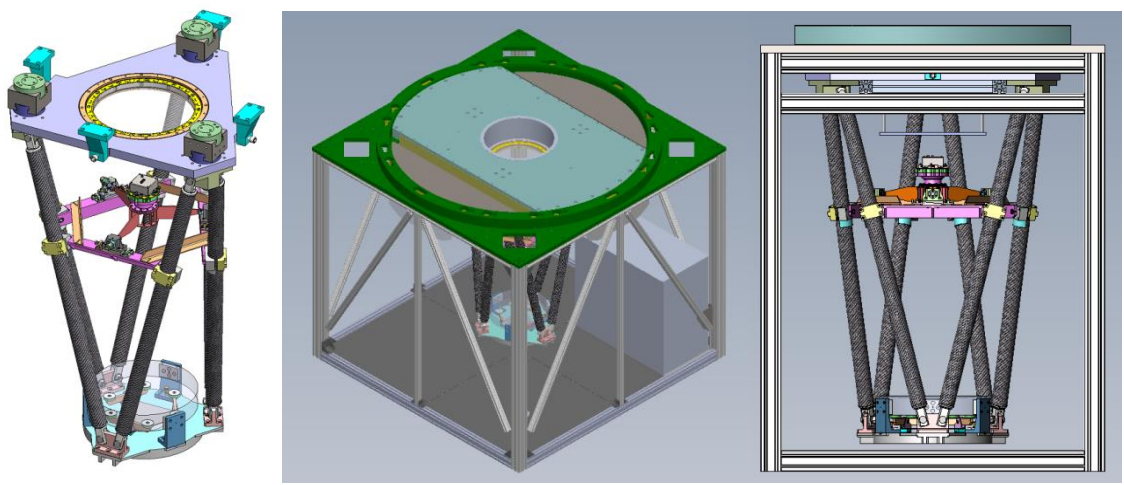

Figure 4. The mechanical supporting structure of the metrology camera (left), the top plate (center) and the Cassegrain box structure (right).

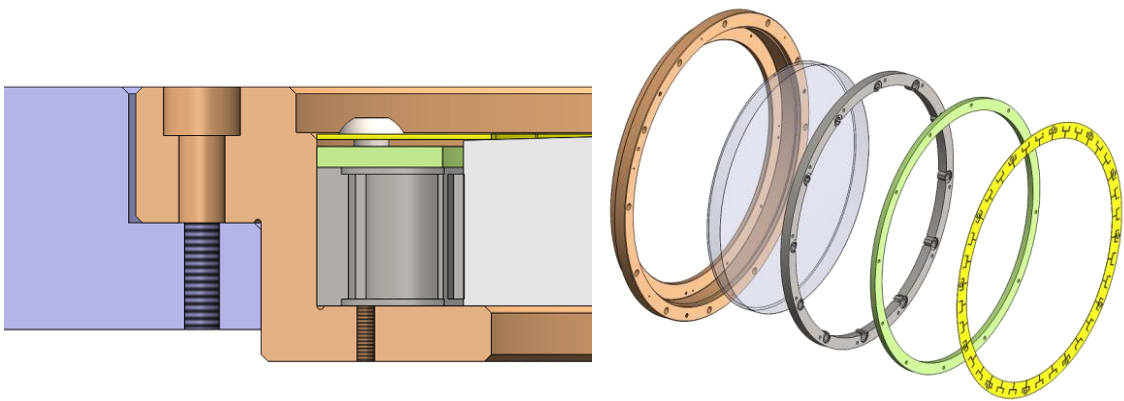

Figure 5 left: Side view of the Schmidt plate holder. From top to bottom are the steel disc spring (yellow), aluminum spacer ring (green), steel roll-pin flexure and steel Schmidt plate holder (brown). Right: The explosive view of the Schmidt plate holder.

The layout of the Schmidt plate mounting structure is shown in figure 5. The Schmidt plate will be installed on a roll-pin flexure[5] to accommodate the thermal expansion/contraction effect. The roll-pin flexure is a stainless steel ring with 12 roll-pins; each pin has 1mm thickness and the pin radius is 0.1mm smaller than the Schmidt plate to provide sufficient preload. Thermal and lens deformation analysis was conducted based on a temperature variation from 20°C to -10°C and elevation angle change from 0° to 90°. The maximum stress on the Schmidt plate at 0° and 45° is 4.29Mpa and 4.12Mpa, respectively. The maximum lens deformation ranges from -0.43µm to 0.37µm normal to the lens surface at 45 degrees. The lens deformation was added to the ZEMAX model for the as-built PSF image simulation.

The spherical mirror mount follows the design used for the mirror mount of the wavefront test system of WFC. Figure 6 is the mechanical layout of the design. The bottom plate is made of invar and weighs about 20kg; it adopts a rib design to reduce weight and plate deformation. The plate surface will be polished to keep a high flatness (20µm/m). Three L-shape mirror side holders are the main supporting structure for the spherical mirror; three guiding grooves are used to maintain the center of the preliminary mirror. The estimated mirror centering accuracy is close to the allowed tolerance of 300µm. The height of the mirror can be adjusted via micrometers and shims under the mirror side holders. The mirror tilt adjusting accuracy is better than 0.02 degrees (alignment method limited, see the next section for details). Six circular pads were used under the spherical mirror to provide auxiliary support; the height of the pads can be fine-tuned by Mitutoyo micrometers. The spherical mirror will be glued on three rectangular pads attaching to the mirror side holders with 3M 2216B/A glue. With a glue area of 1.5cm$^2$ per pad, three pads are strong enough to sustain ~550kg loads, which is more than 20 times higher than the mirror weight. The L-shape mirror side holders and rectangular glue pads are all made of invar to reduce thermal effects.

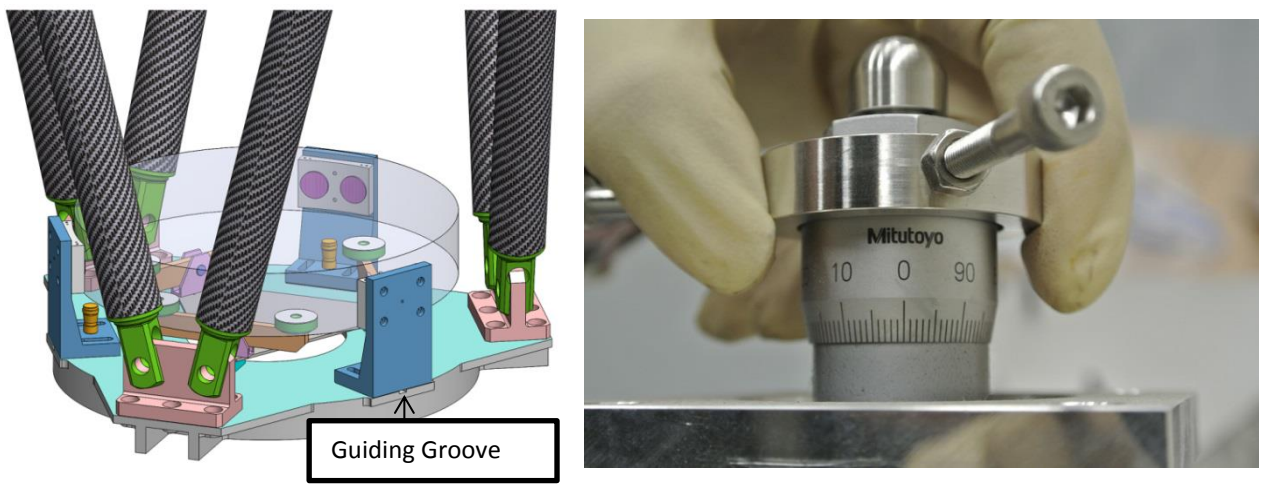

Figure 6. Mechanical layout for the spherical mirror (left). Mitutoyo micrometer used for the mirror bottom supporting pads (right).

The mirror deformation analysis was also conducted with the same temperature and elevation angle conditions as the Schmidt plate. The maximum stresses at 0 and 45 degrees are 0.079MPa and 0.153MPa, respectively. The bottom row shows the mirror deformation normal to the mirror surface. The mirror deformation ranges from -3.13µm to -3.29µm for 0 degrees and -1.25µm to -3.29µm for 45 degrees. All the mirror deformations were added into ZEMAX model for PSF simulation.

The Canon 50M camera and the field flattener will be mounted as a camera module, which has no moving parts involved because of the high strength and low thermal expansion/contraction of the carbon fiber tubes used for the main supporting structure. Figure 7 shows the mechanical layout of the camera module and the supporting structure. To minimize the thermal effects, the shutter and lens cell are all made of invar. The camera housing and camera adapting board are made of aluminum. A kinematic mount is implemented for the Canon 50M camera which allows a quick and accurate camera replacement. Each Canon camera will be pre-adjusted on a corresponding adapting board in the laboratory. The camera can be replaced quickly by removing the screws and relocating a new camera-board set on top of the shutter cell. Two pins were used to secure the mounting accuracy. Another two pins were used for shutter replacement. Since the optics size are small (~ 60mm), the element tilt and decenter within the camera module are determined by the fabrication tolerances.

The camera module sits on a cylindrical structure with three spider veins connecting to the carbon fiber tubes. To adjust the element decenter and tilt of the camera module, an adjusting mechanism was implemented (see figure 8). The adjusting mechanism allows each spider vein to move radially and vertically; fine tuning screws allow an adjusting accuracy of 50μm radially and 20μm vertically.

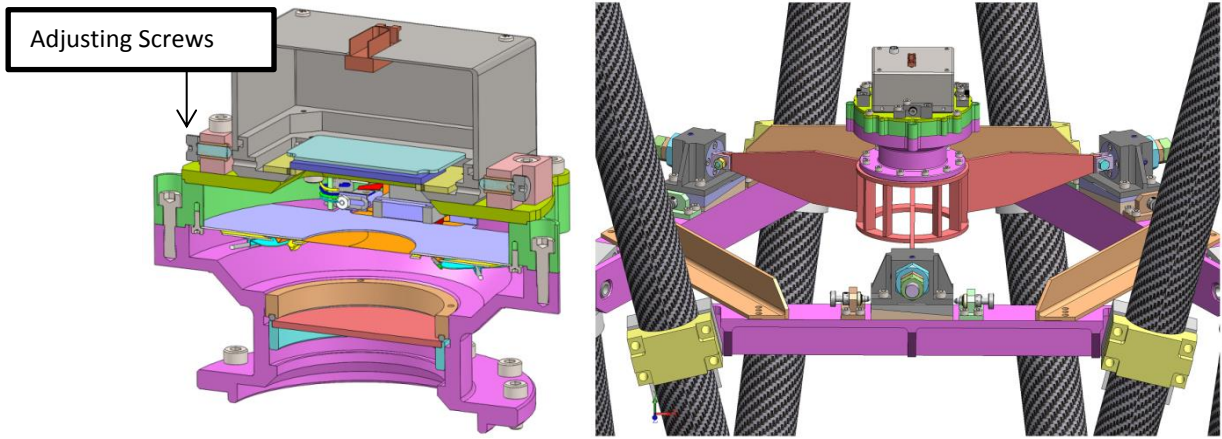

Figure 7. Left panel shows the side view of camera module. From top to bottom are the Canon 50M camera (grey), camera adapting board (green-yellow), shutter cell (green) and lens cell (pink). Right panel shows an overall view of the camera module supporting structure.

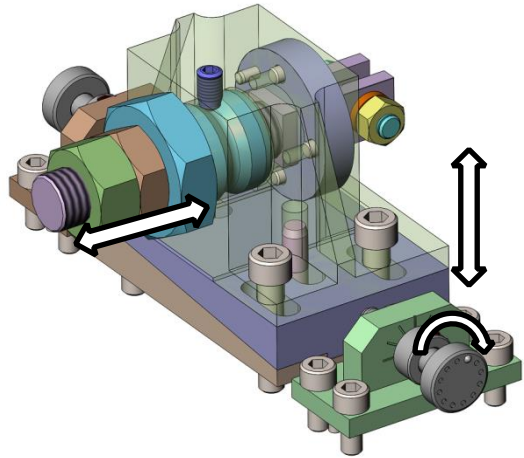

Figure 8. Layout of the camera module adjusting mechanism. Two fine tuning screws on both sides adjust the height of the camera module; the big screw in the center controls the radial position of the camera module.

The thermal and structure analysis was conducted for the metrology camera supporting structure using six carbon fiber tubes (inner diameter 50mm and outer diameter 60mm) to better understand the optical performance change under different observing conditions. For the thermal expansion analysis, the temperature variation was set from -5$^{o}$C to 5$^{o}$C to meet the environmental conditions on Mauna Kea. For the current supporting structure, the largest thermal expansion/contraction is on the order of one micron and can be ignored. For structure deformation analysis, the deformation was derived via ANSYS and added into the ZEMAX design to simulate the as-built PSFs under different elevation angles. With the elongated structure and heavy weight mirror at the bottom of the structure, the maximum shift of the mirror center is about 65μm when the telescope moves from Zenith to the horizontal direction. Such deformation definitely changes the focal position of the metrology camera. To evaluate the image quality and the focal plane position error, we perform a through focus analysis by moving the focal plane from -100μm to 100μm away from the optimized position at Zenith in the ZEMAX model in 10μm steps. Then, we derived the centroid accuracy using PSFs obtained from different focal plane offsets. The results are shown in Figure 9. It is obvious that the metrology camera performance remains stable under different elevation angles, implying the metrology camera requires no focusing mechanism during telescope observations. Given that the acceptable centroid accuracy should be less than 0.02 pixels (shown by the dotted line) the focal plane offset tolerance should be around +/- 40μm away from the focus position. The temperature effect on the metrology camera optics and the supporting structure was also considered in the simulated as-built PSFs; the results show that the temperature changes (20°C to -10°C) will not affect the image centroid estimation accuracy too much.

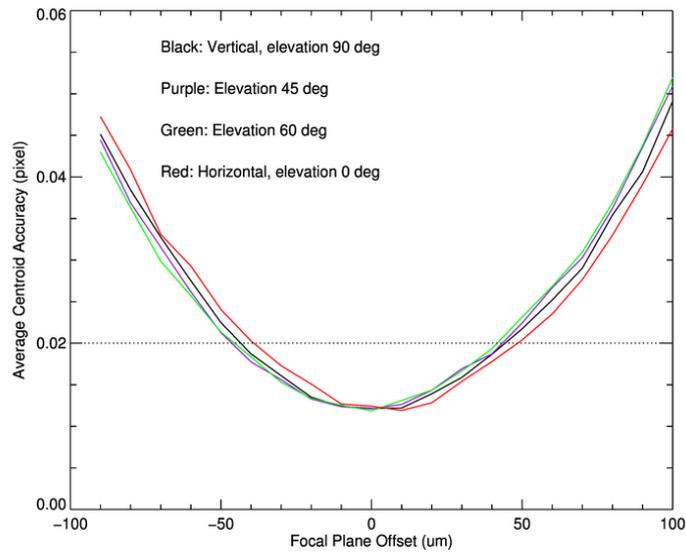

Figure 9. Average centroid accuracy over the entire field of the metrology camera with respect to the focal plane offset under various elevation angles. This plot shows that the metrology camera supporting structure can deliver stable performances under different elevation angles. The horizontal dotted line represents the threshold for the acceptable centroid accuracy, which yields an acceptable focal plane offset of ~ +/- 40µm away from the optimized position.

When installing the assembled metrology camera to the Cassegrain box, the alignment of the metrology camera to the telescope needs to be checked to ensure the FoV covers the focal plane. Based on the distance between the MCS and the PFI and the size of the CMOS sensor, the alignment tolerance is 200µm in centering and 0.01 degrees in tilt. This alignment can be divided into two parts: 1) the alignment of the metrology camera to the Cassegrain box; and 2) the alignment of the Cassegrain box to the telescope. The center of the metrology camera can be adjusted with four screws with an accuracy of 50µm as shown in figure 10. The tilt of the metrology camera can be adjusted with shims with accuracy better than 0.01 degrees. In the second case, the Subaru telescope has provided a Cassegrain box installation repeatability of 150µm (box decentering) and 0.008deg (box tilt), respectively. All these errors satisfy the optically allowed tolerances of 200µm and 0.01 deg. The optically allowed tolerance for the distance between the metrology camera to WFC is 20mm, which can be achieved easily.

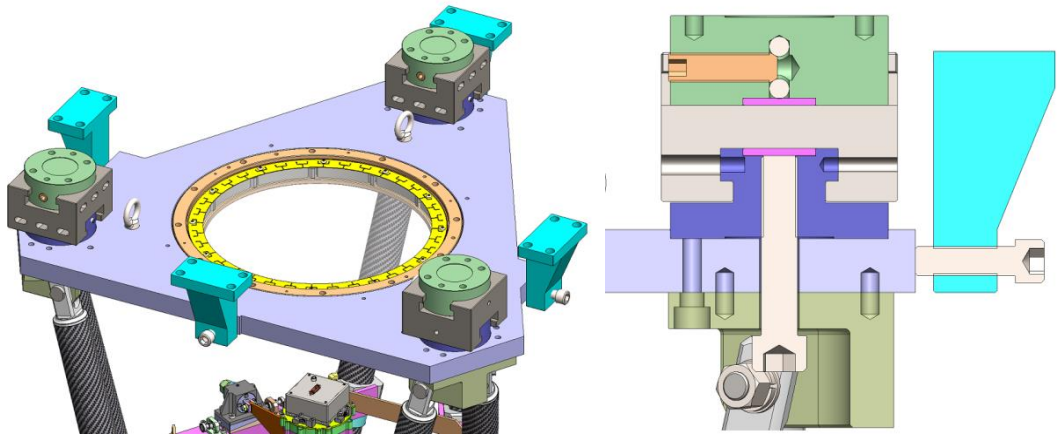

Figure 10. Left panel shows the top view of the three connection mechanisms between the Schmidt plate and the Cassegrain box. Right panel shows the cross section view of one adjusting mechanism. Two screws inside the adjusting mechanism are used to lock the movement of the mechanism; one screw inside the cyan block is used to adjust the centering of the metrology camera.

## 3. METROLOGY CAMERA INTEGRATION & TESTING

Given that all the optics have been fabricated and the mechanical parts are almost ready, the metrology camera integration and testing process will begin soon in summer 2016. The integration and testing will be separated into two stages; the first stage is the integration and alignment in the laboratory in Taiwan; and the second stage will be the commissioning at the observatory.

The integration process in the laboratory starts with the assembly of each optical element: the Schmidt plate, the spherical mirror and the camera module. After each element has been assembled with the associated structure, the alignment process flows as the light pass given the Schmidt supporting plate is the strongest structure for its size and the short distance to the Cassegrain mounting flange. Therefore, the entire integration process will take the Schmidt plate as the reference then flow down to the bottom mirror, and finally to the camera module. Figure 11 shows the flow chart of the metrology camera integration and testing processes in the laboratory. The integration process requires the assistance of a turn table.

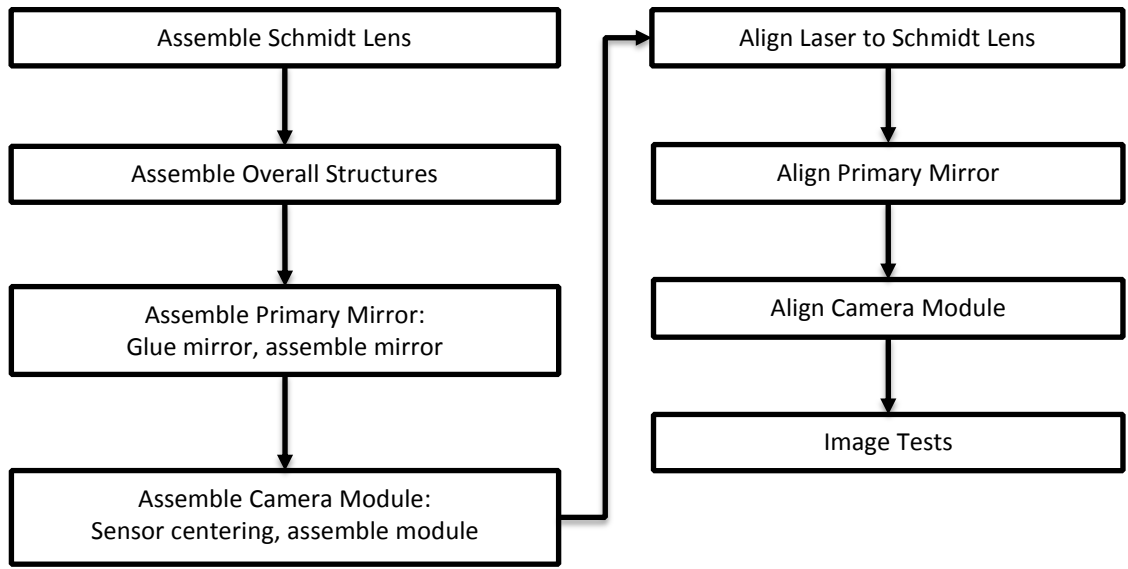

Figure 11. Flow chart of the metrology camera integration and testing procedure in the laboratory.

The first step is to assemble the Schmidt plate. The Schmidt plate will be installed into the roll-pin flexure; then the Schmidt plate lay on the lens holder. The spacer ring and the disc spring will be installed on top of the lens holder. For centering, Nikon has marked a cross at the center of the Schmidt plate, as shown in the left panel of figure 12. To facilitate the lens centering with a laser, a concentric reticle will be used on top of a lens protection cover. Three pins were used to secure the cover installation repeatability. The reticle center will be matched to the marked center on the Schmidt plate with a turn table.

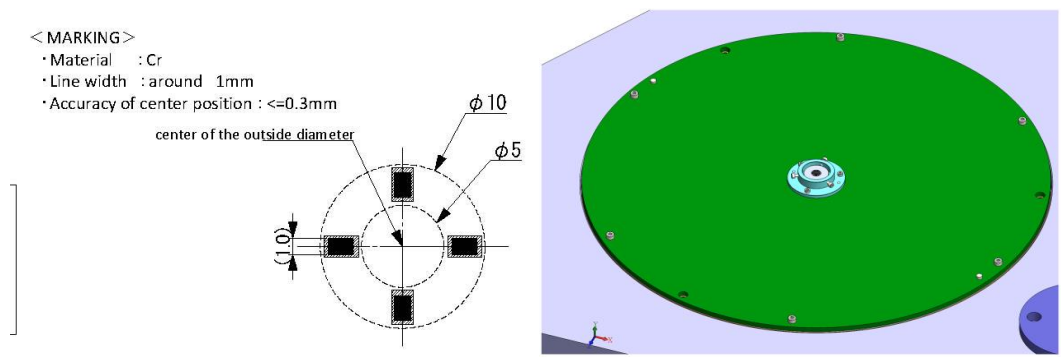

Figure 12. Mark printed on the center of the Schmidt plate (left panel). The Schmidt lens protection cover and the centering reticle are shown in the right panel.

The next step is to assemble the overall supporting structure by connecting the top and the bottom plates using the six carbon fiber tubes. The space between the top and bottom plates will be calibrated with an inner diameter micrometer (Mitutoyo IMZ 139-1500) with an accuracy better than 70µm. The camera module supporting stripes (pink and brown stripes in Figure 9 right panel) will also be installed. The distance between the stripes and the top plate will also be adjusted with the same micrometer to accuracy better than 40µm.

The spherical mirror will be glued on three mounting pads before being installed on the metrology camera main body. The mirror gluing process will be conducted on a flat marble table with an assisting device consisting of two aluminum rings. Figure 13 shows the assisting device which maintains the location of the three mirror mounting pads. Two holes on each mounting pads are used for glue injection and the other two are used for glue thickness adjustment. 100µm shims will be used in keeping the glue thickness to reduce thermal mismatch. After gluing the mirror on the mounting pads, the mirror bottom supporting mechanisms will be assembled and adjusted to the expected height. The spherical mirror will then be laid on the bottom pads and assembled to the mirror side holders. Each mirror side holder will be guided by a groove to a dedicated location. The initial centering accuracy is better than 200µm for the mirror.

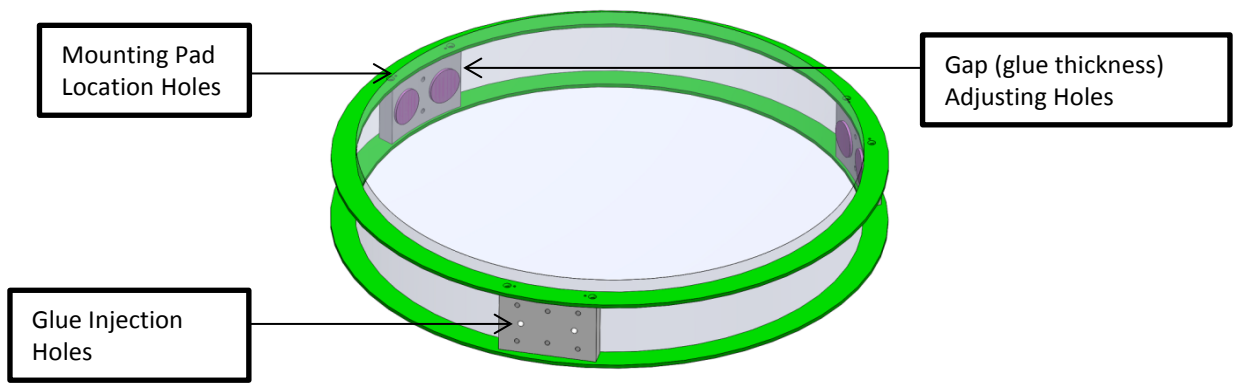

Figure 13. Device used for mirror mounting pads installation.

For the camera module assembling, the camera housing will be attached to the adapting board first; the field flattener and the filter will be pre-adjusted inside the lens cell. The shutter and the shutter cell will be installed on top of the lens cell afterwards. The center of the camera will be matched to the lens cell via the Trioptics turn table during the camera module integration. The assembled camera module will be allocated on the cylindrical structure connecting to the carbon fiber tubes with three radially outward spider veins.

After the assembly of each components the three optics need to be properly aligned. The alignment process requires a laser and a light path greater than 10m to achieve the expected alignment accuracy. The Schmidt plate serves as a reference for the whole alignment process.

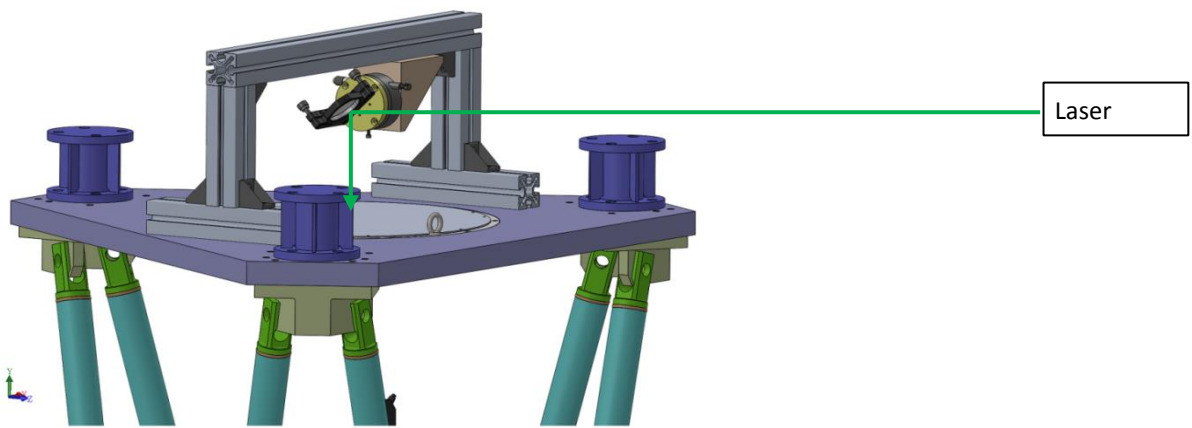

Figure 14. A fold mirror will be set on top of the Schmidt lens holder for the alignment process.

The first step is to set up a flat fold mirror on top of the Schmidt plate and an incident laser beam normal to the plate surface, as shown in figure 14. The laser beam is aimed at the center of Schmidt plate (represented by the reticle center). The light path of 10m guarantees a tilt of Schmidt plate of less than 0.02 degrees, which is 1/3 of the optically allowed value. The second step is to examine the laser beam location on the spherical mirror. Since the mirror surface is spherical, only the mirror tilt will be adjusted until the incident beam coincides with the reflected beam. The mirror tilt can achieve the optically allowed tolerance of 0.02deg with a laser light path of 10m. The last step is to install the camera module and use the PSF image quality to fine tune the element decenter/tilt of the camera module. The justification criteria for claiming a successful alignment are to achieve uniform PSF size across the entire field with a FWHM equal to 10µm. A 14 inch pin-hole mask will be used to simulate the light from the fibers. Figure 15 is the photo of the pinholes mask. On the mask, there are more than 2500 holes with 127µm diameter, which is the size of the fiber cores in PFS. After the confirmation of the image quality, the whole MCS will be tilted to verify the image quality and stability under different elevation angles. The metrology camera will be installed inside a rotary cage and illuminated by a mask located 18m away.

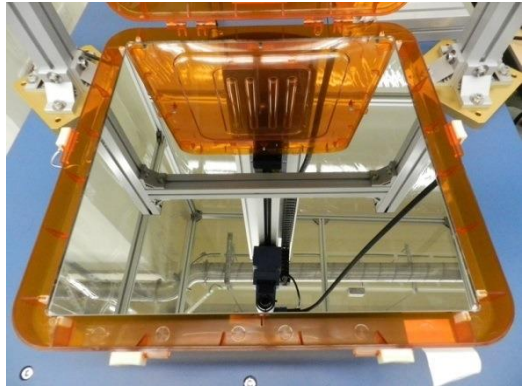

Figure 15. Actual photo of the pinhole mask used for image quality verification by Infinite Graphics Inc.

The metrology camera may need re-alignment after shipping to Hawaii. We will use the same pin-hole mask to check the image quality after the delivery. If needed, the re-alignment process will start from taking off the camera module and setting up the flat fold mirror, then aligning the laser to the Schmidt plate as described in the alignment process above. The next step is to examine the alignment of the spherical mirror via the reflected laser beam. The last step is to re-install the camera module and use the actual PSF image to verify the alignment of the camera module. This re-alignment process requires no assistance of the turn table.

## 4. SUMMARY

The design of the PFS metrology camera is presented. The MCS has passed CDR in Sep 2015 and also a delta CDR in Dec 2015. The design of MCS is proved to be feasible for providing the required functions and accuracy. The optical design has been completed, including detailed performance and tolerance analysis. We have received the CMOS sensor in early 2014 and all the optics of PFS metrology camera were fabricated by end of 2015. We expect to start the integration and testing process in summer 2016 in Taiwan; it will be delivered to Subaru telescope in 2017 for basic function tests and then proceed to the final integration with the rest of PFS.

## ACKNOWLEDGEMENT

We gratefully acknowledge support from the Funding Program for World-Leading Innovative R&D on Science and Technology(FIRST) "Subaru Measurements of Images and Redshifts (SuMIRe)", CSTP, Japan for PFS project. The work in ASIAA, Taiwan is supported by the Academia Sinica of Taiwan.